# A DCCP Congestion Control Mechanism for Wired-cum-Wireless Environments

Ijaz Haider Naqvi
Institut d'Electronique et de Télécommunications de Rennes
INSA Rennes
Rennes, France
ijaz-haider.naqvi@ens.insa-rennes.fr

Tanguy Pérennou
LAAS-CNRS / ENSICA
University of Toulouse
Toulouse, France
tanguy.perennou@ensica.fr

*Abstract*— Existing transport protocols, be it TCP, SCTP or DCCP, do not provide an efficient congestion control mechanism for heterogeneous wired-cum-wireless networks. Solutions involving implicit loss discrimination schemes have been proposed but were never implemented. Appropriate mechanisms can dramatically improve bandwidth usage over the Internet, especially for multimedia transport based on partial reliability. In this paper we have implemented and evaluated a congestion control mechanism that implicitly discriminates congestion and wireless losses in the Datagram Congestion Control protocol (DCCP) Congestion Control Identification (CCID) framework. The new CCID was implemented as a NS-2 module. Comparisons were made with the TCP-like CCID and showed that the bandwidth utilization was improved by more than 30% and up to 50% in significant setups.

## I. INTRODUCTION

The Internet topology is changing fast, and more often than not includes a Wireless Last Hop. Today, packets may get lost on wireless links, for instance due to radio interferences. This is a dramatic change over the assumption that was made in wired networks, where most (if not all) of packet losses were due to network congestion or buffer overflow.

This new situation decreases the performance of classical congestion control mechanisms such as those implemented in TCP or TFRC, which reduce their sending rate each time a packet loss is detected. This systematic decrease principle is also implemented in new generation transport protocols such as SCTP or DCCP. This systematic reduction of the sending rate is useless and inefficient in the case of wireless losses and decreases the per-flow throughput as well as the overall bandwidth utilization.

To improve the situation, a key mechanism is loss discrimination, which determines whether a loss is due to congestion or to a link failure. A number of propositions have been made to perform explicit discrimination, but these propositions make strong assumptions on the existence of specific router properties or specific proxies. Implicit loss discrimination proposals do not rely on such assumptions and need only to be deployed on the end systems.

The present work consists in developing such an implicit loss discrimination mechanism and exploiting it in a congestion control mechanism. The Datagram Congestion Control Protocol (DCCP [1]) implements a framework allowing the inclusion of several congestion control mechanisms. Two mechanisms have been standardized so far: the TCP-like congestion control CCID 2 [2] and TCP-Friendly Rate Control CCID 3 [3].

In this paper we selected an implicit loss discrimination mechanism and implemented it in the NS-2 Network Simulator [5], as a new DCCP congestion control mechanism. The development took advantage of an existing DCCP module for NS-2 [11]. We evaluated the bandwidth utilization and fairness of the mechanism on a wired-and-wireless topology. The wireless link was configured so that it exhibits a bursty loss behavior. The results showed a significant improvement in the performance.

The remainder of the paper is organized as follows: Section II describes related work on loss discrimination in wireless networks; Section III details the implementation of a wireless-aware congestion control mechanism within the NS-2 DCCP implementation; Section IV surveys our simulation results and evaluates the performance of the mechanism; Section V concludes.

## II. RELATED WORK: LOSS DISCRIMINATION

A number of approaches have been studied and suggested for loss discrimination. We can broadly divide loss discrimination in *explicit* loss discrimination and *implicit* loss discrimination.

### A. Explicit Loss Discrimination

In Explicit congestion loss discrimination the sender may be explicitly informed of the type of packet loss by network routers or the end receiver. In these loss discrimination schemes, intermediate network elements (routers) inform the sender about the type of the packet loss as they have got the direct knowledge of the loss type. There are a number of approaches proposed in [4] for explicit loss discrimination. The Split Connection Approach, Link Level Retransmissions and Snoop Protocol [4] are some popular examples.

Supported by HEC Pakistan and LAAS-CNRS / ENSICA.

However, there are cases where link level retransmissions are unsuitable. These include interactive real-time applications, for example VoIP or video-conferencing. Due to their delay sensitivity, this type of applications generally uses UDP instead of TCP. Also explicit loss discrimination schemes can not be used as an end to end protocol and changes must be made at the base stations or intermediate routers to enable them send/receive explicit feedback, which is by no means a simple task. Our work is an effort to have an end to end protocol which at the same time gives us a better performance in a heterogeneous environment and is also equally fair as is TCP.

One of the major problems with UDP is that it has no congestion control to ensure resource fairness. The DCCP [1] addresses this problem by providing congestion control for unreliable transport services. However, this again introduces the discussed problem of inappropriate congestion behavior in the presence of non-congestion losses such as caused by wireless errors. In order to successfully perform congestion control, the reason of a packet loss must be ascertained so that congestion control mechanism is not performed for losses not caused by congestion.

### B. Implicit Loss Discrimination

Implicit loss discrimination schemes do not involve intermediate routers or base stations for detection of loss. There are quite a number of schemes which can be used for loss discrimination.

The Biaz scheme [6, 7] discriminates congestion losses from wireless link losses using Inter-Arrival Times at the receiver. As described in [7], Biaz scheme misclassifies a significant number of congestion losses, preventing the sending rate of a flow from being reduced when the network is congested. A modified version of Biaz (mBiaz) was proposed in [7] which results in lower congestion loss misclassification than the original.

The Spike scheme [7] uses Relative One-way Trip Time (ROTT) to differentiate congestion and wireless losses. The ROTT is used to identify the state of the current connection. If the current state is spike state then losses are assumed to be due to congestion.

The Zig-Zag scheme classifies the losses due to wireless links based on the number of consecutive losses $n$, on the current packet $rott_i$ and the mean $rott_{mean}$ and its standard deviation $rott_{dev}$. A loss is classified as wireless if one among the following conditions is true:

- $n = 1$ and $rott_i < rott_{mean} - rott_{dev}$; or
- $n = 2$ and $rott_i < rott_{mean} - 0.5\ rott_{dev}$; or
- $n = 3$ and $rott_i < rott_{mean}$; or
- $n = 4$ and $rott_i < rott_{mean} + 0.5\ rott_{dev}$;

Otherwise the loss is classified as congestion loss. The mean $rott_{mean}$ and its standard deviation $rott_{dev}$ are calculated using the exponential average:

- $rott_{mean} = (1 - \alpha)\ rott_{mean} + \alpha\ rott_i$
- $rott_{dev} = (1 - 2\alpha)\ rott_{dev} + 2\alpha\ |rott_i - rott_{mean}|$

By definition, ROTT has a high probability of having values greater than ($rott_{mean} - rott_{dev}$): 84% if it were a normalized Gaussian distributed random variable. The threshold of $rott_i > rott_{mean} - rott_{dev}$ intuitively would classify most of the congestion loss correctly. The reasoning behind increasing the threshold with the number of losses encountered is that a more severe loss is associated with higher congestion, and with higher ROTT. This way, a loss event containing four or more packets would be classified as congestion loss only when relatively large ROTT were observed [7].

In most of the implicit loss discrimination schemes, the loss discrimination is performed at the receiver side and the congestion control at the sender side. This calls for the use of *loss notification*, i.e. the process of communicating the results of loss discrimination to the sender side. In [8], an option based scheme is proposed which uses TCP option to inform the sender about the type of packet loss. Of course this adds complexity in a sender side protocol. The use of this loss notification mechanism requires changes to the sender side to implement loss notification and suitable loss adaptation as the loss discrimination is totally receiver based. There are some other schemes which make use of Round trip delays [10] or bandwidth estimation schemes [9] e.g. TCP Westwood for loss discrimination.

We have proposed here a new sender based loss discrimination scheme. We have used the concept of the Zig-Zag scheme for the loss discrimination but it is done at the sender side. All we need for the loss discrimination with Zig-Zag scheme is the value of ROTT, its mean value and its deviated value. The two later can be calculated once you have the value of ROTT. DCCP/TCP-like protocol calculates the value of Round Trip Time (RTT). In the absence of congestion, the time for a packet to reach the receiver must be same as the time for an acknowledgement to reach the sender. This leads to the notion that ROTT is one half of the RTT. So instead of implementing loss discrimination at the receiver side and additional loss notification mechanism, we can implement loss discrimination at the sender side taking ROTT as one half of RTT. This considerably reduces the changes required to be made at the sender side protocol. In our protocol agent, we have implemented loss discrimination at the sender side and have taken ROTT as one half of RTT. The results obtained are not as good as if proper loss notification mechanism would have been implemented but it is a good compromise between simplicity and reasonably good results.

### III. CONGESTION CONTROL IMPLEMENTATION

### A. Principle

We have proposed here a new sender-side loss discrimination scheme derived from the Zig-Zag scheme described in Section II.B. The congestion window (Cwnd) is halved only when a packet loss is detected and identified as a congestion loss. The loss discrimination can be computed only with the current value of ROTT, its mean value and its standard deviation. The two later can be calculated once you have the value of ROTT. In the absence of congestion, the time for a packet to reach the receiver must be the same as the time for an acknowledgement to reach the sender. This leads to approximating the ROTT as one half of the RTT. So instead of implementing loss discrimination at the receiver side and additional loss notification mechanism, we can implement loss discrimination at the sender side taking ROTT as one half of RTT.

Of course if the protocol makes a bad decision then there might be some severe effects. If it wrongly detects a congestion loss to be a wireless link loss then it will not halve its window and it will generate unfairness, i.e. it will take a bigger portion of the bandwidth at the time of congestion. It is quite dangerous and must be avoided. On the other hand if the protocol wrongly detects a wireless link loss to a congestion loss then it will halve its window just like the DCCP TCP-like CCID. In that case the transmission rate will be unnecessarily reduced but it will not affect any other flow and the protocol will remain totally fair. So to misinterpret a wireless link-loss to congestion loss is acceptable but the opposite is not acceptable and must be avoided.

The results obtained are not as good as if proper loss notification mechanism would have been implemented but it is a good compromise between simplicity and reasonably good results as will be shown in Section IV.

### B. Implementation as a NS-2 Agent

The Network Simulator (NS-2) is a discrete event simulator targeted at the network research. It provides support for simulation of different transport layer protocols, routing queue management, trace files and other applications over wired and wireless networks. A DCCP module [11] was implemented in NS-2 [5] and is available as a patch against version 2.29.3. This module implements DCCP CCID2 and CCID3 as NS-2 agents. We used the DCCP/TCP-like Agent implementing CCID2 as a foundation to implement a DCCP/ZigZag Agent, which implements the client-side Zig-Zag scheme for loss discrimination described above.

## IV. EXPERIMENTS AND RESULTS

### A. Experimental Setup

We performed a lot of experiments first without any congestion in the network and then after inserting congestion in the network. As the bursty loss type is the type closest to the reality so we have used the bursty type for most of our experiments. We also studied the performance by changing the number of flows coming towards the bottleneck node. The network topology selected for the experiment is shown in Fig. 1.

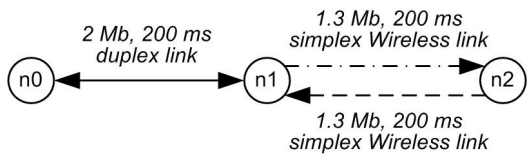

Figure 1. Network topology used in experiments.

It consists of three nodes in series. The first two nodes are connected to each other through a 2Mb bandwidth and 100ms delay duplex wired connection. The second and the third nodes are connected with a wireless simplex link having a bandwidth of 1.3Mb and a propagation delay of 200ms. The source (n0) emits a number of flows towards destination n2 which traverses the bottleneck node n1 in between. For the n0-n1 connection we have used a bi-directional duplex link. For the n1-n2 wireless link we have simulated with two simplex links both of same bandwidth and transmission delay as shown in the figure 1. We have inserted a link loss from n1 to n2 and not from n2 to n1 as we wanted to check the performance of our Agent in a wireless last hop (WLH) scenario. The traffic used is Constant Bit Rate (CBR). For congestion free experiments the accumulated traffic rate of all flows was kept to 1.0Mbps (with bottleneck bandwidth of 1.3Mb). For the experiments that are performed in the presence of the congestion, the accumulated traffic rate of all flows is kept to 1.5Mbps.

To simulate more realistic losses during a wireless link simulation, we must account for *bursts* of packet losses. The most widely known burst error models are the Gilbert Model [12] and the Gilbert-Elliott Model [13, 14]. Markovian models have been developed since [15–20].

We have used a simple 2-state Gilbert Model, with a 0% packet loss rate (PLR) in the "good" state and a 100% PLR in the "bad" state. This 2-state Markovian model is characterized by p, the probability of leaving the good state, and q, the probability of leaving the bad state. Table 1 lists the values of p and q used using during our simulations and the resulting overall PLR, while Fig. 2 shows the comparison of different burst length with the increase in the packet number. In the next experiments, the advertised PLR corresponds to one of these (p, q) couples.

TABLE I. OVERALL PLR FOR DIFFERENT ($P$, $Q$) COUPLES.

| p | q | Packet Loss Rate (PLR) |
|---|---|---|
| 0.001 | 0.6 | 0.176% |
| 0.01 | 0.5 | 1.92% |
| 0.1 | 0.6 | 13.94% |
| 0.1 | 0.4 | 19.8% |

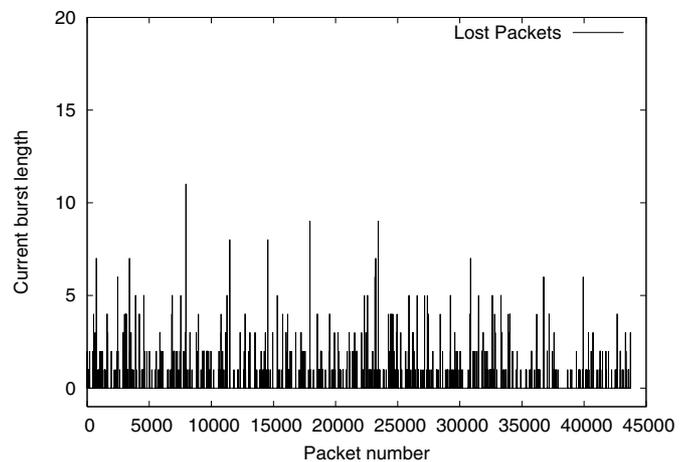

Figure 2. Burst size for different packet numbers using Gilbert Elliot model.

As we can see from Fig. 2 the packets are dropped in bursts and not uniformly.

## B. Results and Analysis

We carried out a lot of experiments on the above topology to test the performance of DCCP with our new congestion control mechanism and we compared the results with the original DCCP/TCPlike protocol agent.

The experiments were done first without generating congestion, with different number of flows and different PLRs. We then increased the rate so that the bandwidth of the bottleneck link becomes lower than the accumulated sum of all flows to check the performance of the algorithm with congestion.

Unless otherwise stated the loss type used is bursty.

## C. Experiments without Congestion

To ensure that there is no congestion in the bottleneck link the accumulated rate of all flows is kept below the bottleneck bandwidth. The bottleneck bandwidth is 1.3Mb and the accumulated rate of all flows is 1.0Mb.

With no loss inserted, the performance of both the protocols is exactly similar. But as we increase the percentage of loss in the wireless link we can see an improvement in the accumulated throughput and mean throughput. Figure 3 shows the comparison of the accumulated throughput with 0.176% loss. We can see from Table 2 that at this loss the mean throughput of both the algorithms is quite low. The reason is that with only one flow the rate of congestion loss misclassification is quite high [7], and every time the protocol assumes a loss as a congestion loss the congestion control mechanism results in a low throughput. We can see with the same loss percentage, the mean throughput increases with multiple flows. The reason is that it is very improbable that every flow experiences congestion at the same time.

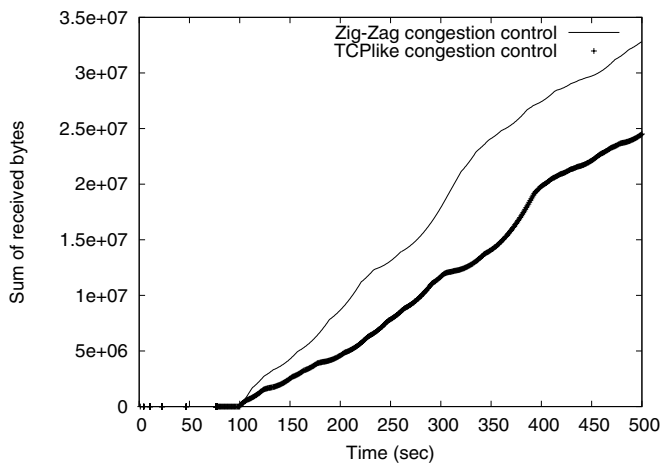

Figure 3.  Figure 3: Accumulated throughputs (0.176% PLR, 1 flow).

There is no improvement in the mean throughput with 0.176% loss and 5 flows. The reason is that the mean throughput with both the algorithms has reached its maximum bandwidth utilization value which is equal to the accumulated rate of all the flows. As the link PLR is not very high and also its is highly improbable that two flows experience a wireless link loss at the same time so we can observe a very good performance with both the Agents and so there is no worth mentioning improvement in the mean throughput. Same is the case with 10 flows.

We can see from the Table 2 that the Zig-Zag scheme gives an improvement in the value of mean throughout obtained after a simulation of 500sec. We have calculated the mean throughput using the last 400 sec as the first 100 seconds are regarded as the warm-up time for the protocol. The warm-up time is the time needed for a protocol to start behaving optimally.

As we increased the PLR the results of our new Agent became better. With one flow obviously the percentage increase was quite significant but the throughput of both the schemes was not very good for the reasons already explained.

TABLE II. PERFORMANCE IN THE ABSENCE OF CONGESTION

| No. Flows | PLR (%) | No. Congestions TCP-like | No. Congestions Zig-Zag | No. Wireless Losses Zig-Zag | Mean Throughput Increase (%) | BW Utilization TCP-like (%) | BW Utilization Zig-Zag (%) |
|---|---|---|---|---|---|---|---|
| 1 | 0.176 | 30 | 23 | 19 | 33.85 | 37.7 | 50.4 |
| 5 | 0.176 | 81 | 81 | 1 | 0.0 | 80 | 80 |
| 1 | 1.92 | 77 | 55 | 45 | 50.48 | 6.03 | 20.18 |
| 5 | 1.92 | 396 | 310 | 172 | 29.94 | 52.7 | 68.5 |
| 10 | 1.92 | 732 | 583 | 159 | 1.86 | 72.2 | 73.55 |
| 10 | 13.94 | 1202 | 1099 | 444 | 21.22 | 20.22 | 24.5 |

With the multiple flows it was quite noticeable that the throughput was a lot better than one single flow. It is interesting to compare the window sizes of the two algorithms when there is some wireless link loss present. As shown in Fig. 4 next page, the window size of DCCP/ZigZag Agent is more than DCCP/TCP-like agent for most of the time because the Zig-Zag algorithm detects some of the wireless losses correctly and does not halve its congestion window in response. In the start of the simulation the window sizes for both of the algorithms is zero. This is the warm-up time, the reason that we have calculated the mean throughputs using the last 400 seconds of the simulation.

For a single flow, Table 2 also shows that while we can achieve an improvement as significant as 50.48%, the throughput and bandwidth utilization remain very low. The mean throughput with the new protocol was 0.2018Mbps when the sending rate was 1Mbps. This confirms that with the only one flow a lot of wireless losses are misclassified as congestion losses [7].

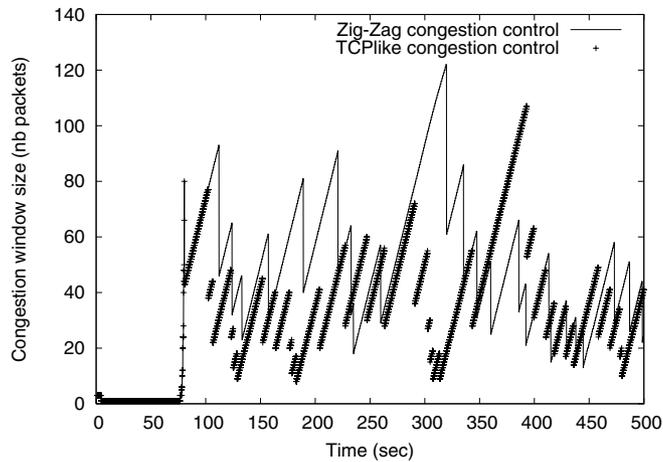

Figure 4. Window sizes of two agents (0.176% PLR, 1 flow).

TABLE III. PERFORMANCE WITH CONGESTIONS

| No. Flows | PLR (%) | No. Congestions TCP-like | No. Congestions Zig-Zag | No. Wireless Losses Zig-Zag | Mean Throughput Increase (%) | BW Utilization TCP-like (%) | BW Utilization Zig-Zag (%) |
|---|---|---|---|---|---|---|---|
| 1 | 0.176 | 30 | 23 | 19 | 33.85 | 33.7 | 50.46 |
| 5 | 0.176 | 102 | 84 | 15 | 0.143 | 96.53 | 96.67 |
| 1 | 1.92 | 77 | 55 | 45 | 50.59 | 10.3 | 15.5 |
| 5 | 1.92 | 396 | 310 | 172 | 30.19 | 52.7 | 68.85 |
| 10 | 1.92 | 717 | 486 | 132 | 5.31 | 87.01 | 91.6 |
| 10 | 13.94 | 1217 | 1075 | 422 | 18.33 | 20.4 | 24.1 |

The mean throughput with the DCCP/TCP-like Agent increases with the increase in the number of flows which is nothing but expected. The reason is that not all the flows will experience wireless errors at the same time. With more flows it is less likely that the wireless losses are synchronized between different flows. The mean throughput of the DCCP/ZigZag agent with one flow and 1.92% PLR is only 0.2018Mbps while with the same PLR and the 5 flows the mean throughput increases to 0.8524Mbps. As shown in the Table 2, the numbers of congestion losses of both the agents have also increased significantly with the number of flows. The percentage increase in the mean throughput is 29.94% with five flows as compared to 50.48% with one flow.

If we further increase the PLR up to 13.94% (which is highly improbable in the real scenarios) the two algorithms have a very low throughput with 1 flow. But if we also increase the number of flows the throughput starts increasing. At this high PLR we got 23.4% improvement with 5 flows and 21.22% improvement with 10 flows.

To sum up on experiments without congestion, it can be noticed that the mean throughputs achieved during the experiments are quite low when there is only 1 flow even in the absence of congestion. The reason is that Zig-Zag scheme is not very good in classifying wireless and congestion losses with one flow. But we have seen throughputs as high as 0.8524 Mbps and 0.9562 Mbps with 5 and 10 flows respectively. We achieved up to 95% bandwidth utilization even in the presence of wireless loss.

D. *Experiments with Congestion*

We also performed some experiments when there was congestion in the network: for these experiments the bottleneck bandwidth was 86% of the accumulated traffic rate of all the flows. The bottleneck bandwidth was kept 1.3Mb but the accumulated transmission rate of all flows was increased to 1.5Mbps.

For 1 flow the results were very similar to the ones without congestion. Even the numbers of congestion and wireless losses detected were quite similar as can be seen from Table 3 and Table 2.

As the number of flows increases, so does the number of congestion losses. The misclassification rate of the Zig-Zag scheme starts decreasing [7]. This results in a better performance in the presence of congestion.

As shown in Table 3 we got 30.19% increase in the mean throughput in the presence of congestion with 5 flows and 1.92% link loss while with the same set of parameter the increase was 29.94% without congestion. When we further increase the number of flows the mean throughput increases for the two algorithms and we see very high bandwidth utilization. For example with 10 flows and 1.92% link loss the bandwidth utilization for DCCP/ZigZag is 91.6% where as for the same parameters the bandwidth utilization for TCP-like is 87%. So we have 4.6% increase in bandwidth utilization where as the increase in the throughput is 5.31%. When we increase the PLR further, the percentage increase in the throughput increases as the number of congestion losses becomes very high at 13.94% PLR as shown in Table 3.

There is no significant increase in the mean throughput when the experiments are performed with 1.92% link loss and 10 flows. In the presence of congestion the increase is 5.31% (from Table 3) and it reduces to 1.86% in the absence of congestion (Table 2). On the other hand we get significant improvement with the same PLR when the number of flows are 5 and 1 i.e. approximately 30 and 50% respectively. The reason for a moderate improvement in the mean throughput is that with the increase in flows throughput of the DCCP/TCP-like Agent also increases as it is highly improbable to get a congestion loss at the same time for all flows. So with the increase in the throughput the percentage increase decreases. However if we look at the number of congestion and wireless losses, there are significant number of wireless losses which are detected correctly (156) and thus preventing the protocol to execute its congestion control mechanism.

As shown in the Table 3 the percentage increase in the throughput at 13.94% loss in the presence of congestion is 18.33% as compared to 21.224% in the absence of congestion. The bandwidth utilization at this high loss is very poor i.e. 24.14% for DCCP/ZigZag and 20.4% for DCCP/TCP-like Agent. The reason for the decrease in the improvement is that high number of congestion losses has affected the throughput achieved.

### E. Experiments with Uniform Losses

We also performed a few experiments with uniformly distributed losses but found out that the performance with both algorithms is not that good. Table 4 shows the performance of the two algorithms with or without congestion inserted in the network. The percentage increase with or without congestion has decreased for the same number of flows and PLR. Also the bandwidth efficiency has also decreased considerably. This also ascertains the notion that the loss model has an important impact on the interpretation of results.

TABLE IV. PERFORMANCE WITH UNIFORM LOSSES

| No. Flows | PLR (%) | No. Congestions TCP-like | No. Congestions Zig-Zag | No. Wireless Losses Zig-Zag | Mean Throughput Increase (%) | BW Utilization TCP-like (%) | BW Utilization Zig-Zag (%) |
|---|---|---|---|---|---|---|---|
| 5 (without congestion) | 1.92 | 464 | 396 | 178 | 24.2 | 42 | 52.2 |
| 5 (with congestion) | 1.92 | 464 | 405 | 152 | 22.4 | 42 | 51.4 |

## V. CONCLUSION

The addition of a wireless loss detection capability has given DCCP a new dynamics and has made it more powerful than it was before. We studied a modified DCCP for the transmission of data in WLH scenario and observed that with the help of simple modifications, it achieves better performance than the regular TCP-like congestion control mechanism. The Zig-Zag algorithm for the discrimination of wireless losses was implemented on the sender side (which eliminated the need for loss notification) and we achieved up to 50% improvement in the mean throughput. The results show that even in the presence of wireless link loss we were able to achieve the bandwidth utilization ratio as good as 90%.

These results can still be improved by reducing the misclassification rate of the loss discrimination mechanism. This can be done either by investigating more on the Zig-Zag scheme, e.g. implementing it on the receiver side, or by implementing alternate schemes such as the mBiaz scheme. These schemes can also be implemented on top of the DCCP CCID3 TFRC rather than on top of DCCP CCID2 TCP-like as was done in this paper, so that improvements in bandwidth utilization may be exploited by flows needing smoother changes in the sending rate.

Finally, the most appropriate schemes may be implemented in a real DCCP stack, such as the Linux 2.6.14+ implementation [20], and tested either on a wireless emulation platform such as W-NINE [21], or on a real wired-cum-wireless environment.

### ACKNOWLEDGEMENT

This work was supported in part by the HEC Pakistan and LAAS-CNRS / ENSICA, France. It was done during Ijaz Haider Naqvi during his MS internship in LAAS-CNRS / ENSICA.